\documentclass[twocolumn,showpacs,prl,amsmath,amssymb,superscriptaddress]{revtex4}

\usepackage{graphicx}
\usepackage{dcolumn}
\usepackage{bm}
\usepackage{amssymb}

\begin{document}

\title{Ultrasensitive Proximity Josephson Sensor with Kinetic Inductance Read-Out}

\author{F. Giazotto}
\email{giazotto@sns.it}
\affiliation{NEST CNR-INFM and Scuola Normale Superiore, I-56126 Pisa, Italy}
\author{T. T. Heikkil\"a}
\affiliation{Low Temperature Laboratory, Helsinki University of
Technology, P.O. Box 3500, FIN-02015 TKK, Finland}
\author{G. Pepe}
\affiliation{CNR-INFM Coherentia and Dipartimento Scienze Fisiche, Universit\`a di Napoli "Federico II," Monte Sant'Angelo, I-80125 Napoli, Italy}
\author{P. Helist\"o}
\affiliation{VTT Information Technology, Tietotie 3, Fin-02150 Espoo, Finland}
\author{A. Luukanen}
\affiliation{Millilab, VTT, Tietotie 3, Fin-02150 Espoo, Finland}
\author{J. P. Pekola}
\affiliation{Low Temperature Laboratory, Helsinki University of
Technology, P.O. Box 3500, FIN-02015 TKK, Finland}

\begin{abstract}
We propose a mesoscopic kinetic-inductance radiation detector based
on a long superconductor--normal metal--superconductor Josephson
junction. The operation of this proximity Josephson sensor (PJS)
relies on large kinetic inductance variations under irradiation due
to the exponential temperature dependence of the critical current.
Coupled with a dc SQUID readout, the PJS is able to provide a signal
to noise (S/N) ratio up to $\sim 10^3$ in the THz regime if operated
as calorimeter, while electrical noise equivalent power (NEP) as low
as $\sim 7\times10^{-20}$ W/$\sqrt{\text{Hz}}$ at 200 mK can be
achieved in the bolometer operation. The high performance together
with the ease of fabrication make this structure attractive as an
ultrasensitive cryogenic detector of THz electromagnetic radiation.

\end{abstract}


\maketitle

Superconducting single-photon detectors \cite{Kerman,Goltsman,Houck} offer
high infrared detection efficiency, high-speed timing resolution and
few-nanosecond reset times. They have been applied in several fields
including spectroscopy of ultrafast quantum phenomena \cite{Jaspan},
optical communications \cite{Robinson}, quantum cryptography
\cite{qcryptography}, and fast digital circuit testing
\cite{Korneev}. On the other hand, a wide potential for
superconducting nanoscale detectors used as advanced bolometers is
also expected in several astrophysical space applications, where
bolometers are promising candidates to meet future needs of cooled
telescopes. The interest lies in the negligible Johnson noise they
show with a NEP as low as $10^{-18}$
W/$\sqrt{\text{Hz}}$. Hot-electron resistive
microbolometers and kinetic inductance superconducting detectors
(KIDs) represent high performance devices able to reach NEP$\lesssim
10^{-19}$ W/$\sqrt{\text{Hz}}$ at $T\geq 1$ K \cite{Sergeev}. KIDs
\cite{Grossman} offer about the same NEP and response
time as resistive bolometers and hot electron detectors, and they
can operate at temperatures much below the critical
temperature where the generation-recombination noise is small
thanks to the reduced number of quasiparticles.

Here we propose a KID based on a long
superconductor-normal metal-superconductor Josephson junction. It
exploits large kinetic inductance variations under irradiation
thanks to the exponential temperature dependence of the
supercurrent, and yields a high S/N ratio ($\sim 10^3$ around 40
THz) and a low NEP ($\sim 7\times 10^{-20}$ W/$\sqrt{\text{Hz}}$ at
200~mK). The ease of implementation combined with large array
scalability make this structure promising as a sub-Kelvin
ultrasensitive detector of far- and mid-infrared electromagnetic radiation.

\begin{figure}[t!]
\includegraphics[width=\columnwidth,clip]{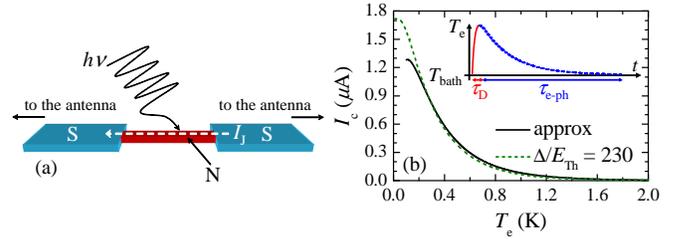}
\caption{(color online) (a) Scheme of the PJS. Incident
electromagnetic radiation elevates the electron temperature
($T_{\text{e}}$) in the N wire, thus strongly suppressing
the Josephson current. This leads to a large enhancement of the
junction kinetic inductance. (b) Supercurrent $I_{\text{c}}$ vs
electron temperature $T_{\text{e}}$ of a long SNS Josephson junction
calculated at $\phi =\pi/2$ from Eq. (\ref{supercurrent}) (full
line), and for $\Delta/E_{\text{Th}}=230$ (dashed line). The inset
shows a sketch of time evolution of $T_{\text{e}}$ after absorption
of radiation (see text).
}
\label{fig1}
\end{figure}

The structure we envision is sketched in Fig. 1(a) and
consists of a diffusive normal metal (N) wire of length $l$ coupled
to two superconducting leads (S) through transparent contacts, thus
realizing a SNS Josephson weak link. An antenna eventually couples
the incident radiation to the N wire. We assume that the Josephson
junction is long, i.e., $\Delta \gg \hbar D/l^2=E_{\text{Th}}$,
where $\Delta$ is the S gap, $D$ is the diffusion
coefficient of N, and $E_{\text{Th}}$ is the Thouless
energy. The radiation coupled to the junction heats the electrons
in N to temperature $T_{\text{e}}$. For
$E_{\text{Th}} \ll k_B T_{\text{e}} \ll \Delta$, the Josephson
current is $I_{\text{J}}=I_{\text{c}}\text{sin}(\phi)$, where
$\phi$ is the phase difference across superconductors, and
\cite{SNS}
\begin{equation}
I_{\text{c}}=\frac{64\pi k_{\text{B}}T_{\text{e}}}{(3+2\sqrt 2)eR_{\text{N}}}\sqrt{\frac{2\pi k_{\text{B}}T_{\text{e}}}{E_{\text{Th}}}}\text{exp}\left(- \sqrt{\frac{2\pi k_{\text{B}}T_{\text{e}}}{E_{\text{Th}}}}\right)
\label{supercurrent}
\end{equation}
is the junction critical current. Hence, in this limit,
$I_{\text{c}}$ depends exponentially on the electron temperature and
is independent of the phonon temperature $T_{\text{bath}}$. In
Eq.~\eqref{supercurrent}, $R_\text{N}=\rho l/\mathcal{A}$ is the
normal-state resistance of the junction,
$\rho=(\nu_{\text{F}}e^2D)^{-1}$ is the wire resistivity,
$\mathcal{A}$ its cross section, and $\nu_{\text{F}}$ is the density
of states at the Fermi level in N. For our simulation
 we choose a 10-nm-thick silver (Ag) wire with
$l=1\mu$m, width of 100 nm (volume $\Omega=10^{-21}$ m$^3$),
$\nu_{\text{F}}=1.0\times 10^{47}$ J$^{-1}$m$^{-3}$, and $D=0.01$
m$^2$s$^{-1}$ . With the aforementioned parameters
$R_{\text{N}}\simeq 38\,\Omega$, and $E_{\text{Th}}\simeq
6.6\,\mu$eV. By choosing, for instance, Nb as S electrodes
($\Delta=1.52$ meV) we get $\Delta/E_{\text{Th}}\simeq 230$, thus
providing the frame of the \emph{long} junction limit. The critical current
$I_{\text{c}}$ vs $T_{\text{e}}$ is shown in
Fig. 1(b) for $\phi =\pi/2$ and $\Delta/E_{\text{Th}}=230$ (dashed
line). In our case, $I_{\text{c}}$ saturates around
$1.7\,\mu$A at $T_{\text{e}}\simeq 50$ mK, and is
suppressed by a factor of $\sim 20$ at 1 K due to the
exponential dependence on $T_{\text{e}}$. For a comparison, the
approximated result from Eq. (\ref{supercurrent}) is also shown
(full line), and will be used in the following. Such
 suppression of $I_{\text{c}}$
produces a large enhancement of the junction kinetic inductance
($L_{\text{k}}$), defined as
$L_{\text{k}}=\hbar/(2eI_{\text{c}})$. As we shall show,
measuring $L_{\text{k}}$ variations with a suitable readout scheme
allows to accurately detect the radiation absorbed by the SNS
junction.

In order to understand the operation principle of the PJS as a
calorimeter (i.e., in pulsed excitation operation) as well as a
bolometer (i.e., in continuous excitation operation) it is useful to
consider the inset of Fig. 1(b), which shows a sketch of time
evolution of $T_{\text{e}}$ in N after the
arrival of a photon. We assume that $T_{\text{e}}$ is
elevated with respect to $T_{\text{bath}}$, depending on the energy
of the impinging photon and uniformly along N, over a time
scale set by the diffusion time $\tau_{\text{D}}=l^2/D$ [see the red
line in the inset of Fig. 1(b)]. With our parameters
$\tau_{\text{D}}=10^{-10}$ s. Then, after the absorption of a
photon, $T_{\text{e}}$ relaxes toward $T_{\text{bath}}$
over a time scale set by the electron-phonon interaction
time ($\tau_{\text{e-ph}}$), given by
$\tau_{\text{e-ph}}=1/(\alpha T_{\text{bath}}^3)$ \cite{RMP}, where
$\alpha\approx 0.34\Sigma/(k_{\text{B}}^2\nu_{\text{F}})$,
and $\Sigma$ is the
electron-phonon coupling constant. By setting $\Sigma=5\times 10^8$
Wm$^{-3}$K$^{-5}$, as appropriate for Ag \cite{RMP},
$\tau_{\text{e-ph}} \sim 1\times
10^{-4}\ldots 1\times 10^{-7}$ s in the $0.1\ldots 1.0$ K
temperature range, so that $\tau_{\text{e-ph}}\gg \tau_{\text{D}}$
[see the blue line in the inset of Fig. 1(b)]. In the following we
analyze both the pulsed
and the continuous excitation mode of the PJS.

In the pulsed mode, after the arrival of a photon of frequency $\nu$
at time $t=0$, the electron temperature in N can be
determined by solving the heat equation $C_{\text{e}}(\partial
T_{\text{e}}/\partial t)=P_{\text{opt}}$ \cite{RMP}, where
$C_{\text{e}}=(\pi^2\nu_{\text{F}}k_{\text{B}}^2T_{\text{e}})/3$ is
the electron heat capacity, and $P_{\text{opt}}=(2\pi
\hbar\nu/\Omega)\delta(t)$ is the optical input power per volume per
incident photon. In writing the
heat equation we neglected the spatial dependence of $T_{\text{e}}$
in N, as well as the interaction with the lattice phonons,
the latter occurring on a time scale $\tau_{\text{e-ph}}\gg
\tau_{\text{D}}$. From the solution of the heat equation we
get $T_{\text{e}}(\nu)=\sqrt{T_{\text{bath}}^2+12\hbar\nu/(\pi
\Omega\nu_{\text{F}}k_{\text{B}}^2)}$, which shows that
small N volumes are required to achieve large
enhancement of $T_{\text{e}}$. This
condition can be easily met in metallic SNS junctions, where N
island with volumes below $10^{-21}$ m$^3$ can be routinely
fabricated with the present technology.
\begin{figure}[t!]
\includegraphics[width=\columnwidth,clip]{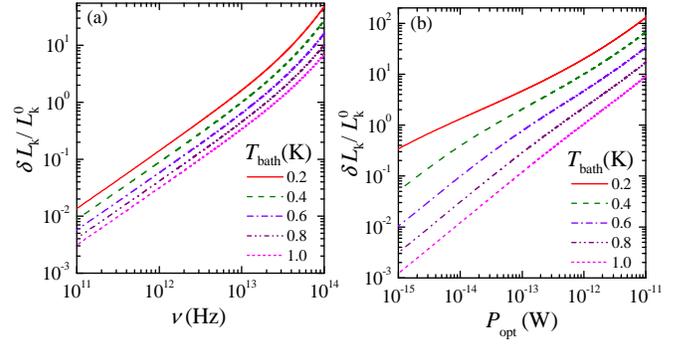}
\caption{(color online) (a) $\delta L_{\text{k}}/L_{\text{k}}^0$ vs $\nu$ at different $T_{\text{bath}}$.
(b) $\delta L_{\text{k}}/L_{\text{k}}^0$ vs $P_{\text{opt}}$ calculated at different $T_{\text{bath}}$.
}
\label{tS}
\end{figure}
The relative variation of the kinetic inductance, i.e., the quantity
$\delta
L_{\text{k}}/L_{\text{k}}^0=[L_{\text{k}}(\nu)-L_{\text{k}}(0)]/L_{\text{k}}(0)$
is displayed in Fig. 2(a) as a function of $\nu$ at different
$T_{\text{bath}}$. In the present structure, $\delta
L_{\text{k}}/L_{\text{k}}^0$ of about $14\%$ for 1 THz photon, and
around $163\%$ for 10 THz photon can be achieved at $200$ mK. At
higher bath temperatures $\delta L_{\text{k}}/L_{\text{k}}^0$ is
reduced, and obtains values of about $3\%$ at 1 THz and around
$35\%$ at 10 THz at $T_{\text{bath}}=1$ K. Such kinetic inductance
variations allow for a very large signal to noise ratio for
single-photon detection.

The PJS operation in continuous excitation
can be described by considering those mechanisms which drive power
into the N electrons. At low temperature (typically below 1 K), the
main contribution in metals is related to electron--phonon heat
flux which can be modeled by
$\dot{Q}_{\text{e-ph}}=\Sigma\Omega(T_{\text{e}}^5-T_{\text{bath}}^5)$
\cite{RMP}. The steady-state $T_{\text{e}}$ under
irradiation with a continuous power $P_{\text{opt}}$ thus follows
directly from the solution of the energy balance equation
$P_{\text{opt}}+\dot{Q}_{\text{e-ph}}=0$, which gives
$T_{\text{e}}(P_{\text{opt}})=\sqrt[5]{(P_{\text{opt}}/\Sigma\Omega)+T_{\text{bath}}^5}$.
This expression shows that both reduced $\Omega$ and small $\Sigma$
are required to maximize $T_{\text{e}}$ enhancement upon power
irradiation. The impact of continuous power excitation on the
junction kinetic inductance is shown in Fig.~\ref{tS}(b) which shows
$\delta
L_{\text{k}}/L_{\text{k}}^0=[L_{\text{k}}(P_{\text{opt}})-L_{\text{k}}(0)]/L_{\text{k}}(0)$
versus $P_{\text{opt}}$ at several $T_{\text{bath}}$. Notably,
$\delta L_{\text{k}}/L_{\text{k}}^0$ as large as $\simeq 130\%$ for
$P_{\text{opt}}=10$ fW and $\simeq 2000\%$ for $P_{\text{opt}}=1$ pW
at $T_{\text{bath}}=0.2$ K can be achieved. At higher bath
temperatures $\delta L_{\text{k}}/L_{\text{k}}^0$ gets reduced, reaching
values $\simeq 1\%$ for $P_{\text{opt}}=10$ fW and $\simeq 100\%$
for $P_{\text{opt}}=1$ pW at 1 K.

\begin{figure}[t!]
\includegraphics[width=\columnwidth,clip]{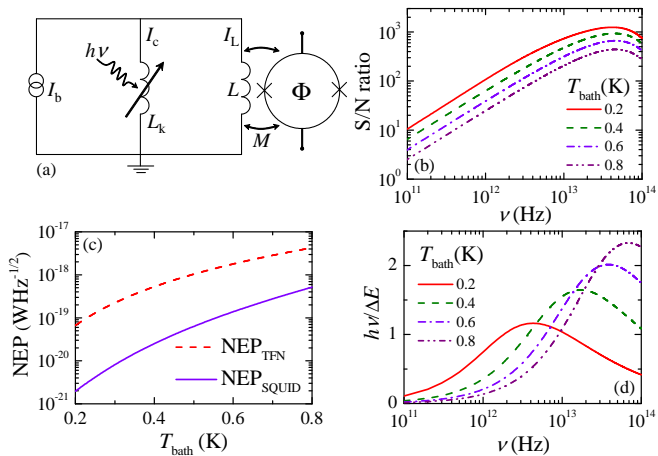}
\caption{(color online) (a) Scheme of the KID with a dc SQUID readout. (b) S/N ratio vs $\nu$ calculated at different $T_{\text{bath}}$. (c) Temperature dependencies of NEP due to thermal fluctuation noise (TFN) and to the SQUID readout.
(d) Resolving power vs $\nu$ calculated at different $T_{\text{bath}}$.
}
\label{tF}
\end{figure}
We now turn on discussing the PJS performance by considering a
superconducting quantum interference device (SQUID) readout, as
shown in Fig. 3(a) \cite{multiplexing}. A constant bias current
$I_{\text{b}}$ divides into two parts, i.e., one flowing through the
SNS junction \cite{sinusoidalnote}, and the other ($I_{\text{L}}$)
through a load inductor ($L$) coupled to a dc SQUID. Upon
irradiation, an enhancement of $L_{\text{k}}$ results in a variation
of $I_{\text{L}}$, thus producing a magnetic field which is detected
by the SQUID.
The magnetic flux generated by the incident radiation is given by
$\Phi=M I_{\text{L}}$, where $M$ is the mutual inductance between
the SQUID and the SNS junction loop. In the linearized regime, i.e.,
by assuming $LI_{\text{L}}\ll \Phi_0$ where $\Phi_0$ is the flux
quantum, we get $I_{\text{L}}\approx
I_{\text{b}}\Phi_0/(\Phi_0+LI_{\text{c}})$, and
$dI_{\text{L}}/dI_{\text{c}}\approx
LI_{\text{b}}\Phi_0/(\Phi_0+LI_{\text{c}})^2$. In the pulsed
detection mode, the signal to noise ratio (S/N) can be readily
expressed as
\begin{equation}
\frac{\text{S}}{\text{N}}=\frac{(d\Phi/dT_{\text{e}})\delta
T_{\text{e}}}{\delta
\Phi_{\text{n}}\sqrt{\omega}}=\left|\frac{M(dI_{\text{L}}/dI_{\text{c}})(dI_{\text{c}}/dT_{\text{e}})\delta
T_{\text{e}}}{\delta \Phi_{\text{n}}\sqrt{\omega}}\right|,
\label{S/N}
\end{equation}
where $\delta \Phi_{\text{n}}$ is the flux sensitivity of the dc
SQUID, and $\omega$ its bandwidth. The S/N ratio versus $\nu$ is
shown in Fig. 3(b) at different $T_{\text{bath}}$. Here we set
$L=100$ nH, $M=10$ nH, $\omega =1$ MHz,
$\delta\Phi_{\text{n}}=10^{-7}\Phi_0/\sqrt{\text{Hz}}$ \cite{SQUID},
and $I_{\text{b}}=0.8I_{\text{c}}(\nu)$. Notably, very high S/N
ratios can be achieved with the PJS in the 100 GHz - 100 THz
frequency range. The S/N ratio is maximized around 40 THz where it
obtains values $\sim 1.2\times10^3$ at $T_{\text{bath}}=0.2$ K.
 In the bolometer operation, on the other
hand, an important figure of merit is the NEP, which is due to
several uncorrelated noise sources. In our case, the dominant
contribution is due to thermal fluctuation noise-limited NEP
(NEP$_{\text{TFN}}$), given by
NEP$_{\text{TFN}}=\sqrt{5k_{\text{B}}\Sigma\Omega(T^6_{\text{e}}+T^6_{\text{bath}})}$
\cite{RMP}, while the contribution due to Johnson noise is absent,
thanks to the operation of the junction in the dissipationless
regime. The contribution of the SQUID readout to NEP
(NEP$_{\text{SQUID}}$) can be determined by setting $\text{S/N}=1$,
$\omega=1$ Hz, and solving Eq.~\eqref{S/N} for $P_{\text{opt}}$.
Figure 3(c) shows the NEP$_{\text{TFN}}$ (dashed line) and
NEP$_{\text{SQUID}}$ (full line) vs $T_{\text{bath}}$.
NEP$_{\text{SQUID}}$ is significantly smaller than
NEP$_{\text{TFN}}$, and the latter can be as low as $\simeq
7\times10^{-20}$ W/$\sqrt{\text{Hz}}$ at 0.2 K. Further reduction of
NEP$_{\text{TFN}}$ is possible by lowering $\Omega$ as well as by
exploiting materials with lower $\Sigma$. Above, we have discussed
the electrical NEP --- the optical NEP is of the same order of
magnitude. This is because the resistance of the device can be
easily matched to common broadband self-similar lithographic
antennas. The PJS resolving power ($2\pi \hbar \nu/\Delta E$) vs
frequency, where $\Delta E\approx
2\sqrt{2\text{ln}2}\text{NEP}_{\text{TFN}}(\nu)\sqrt{\tau_{\text{e-ph}}}$
is the energy resolution of full width at half maximum \cite{RMP},
is displayed in Fig. 3(d) for different $T_{\text{bath}}$. In
particular, the figure shows that resolving power values between
$\sim 1.2$ and $\sim 2.3$ can be achieved in the $5\ldots70$ THz
frequency range for $T_{\text{bath}}\gtrsim 400$ mK, thus making the
PJS suitable for \textit{far}- and \textit{mid-infrared}
single-photon detection.

The mechanism for the supercurrent in SNS junctions is due to the
proximity effect, giving rise in the N local density of states to an
energy minigap of size $E_g=c(\phi)E_{\rm Th}$ with $c(0) \approx
3.1$ \cite{zhou98}, which we have ignored in the expressions for
the heat capacity and electron-phonon coupling. Due to the minigap,
both of these quantities are reduced inside the N wire, further
improving the device resolution. Moreover, the density of states is
not divergent at the minigap edge, contrary to bulk superconductors,
and thereby the generation-recombination noise is reduced. For the
chosen parameters the minigap $E_g\sim h \cdot 5$ GHz, and thereby
radiation from low frequencies $\nu < E_g/h$, which typically is not
part of the measured signal, should not couple in the device.

In summary, we have analyzed a proximity Josephson sensor (PJS)
based on a long SNS junction in the kinetic inductance
mode. Signal-to-noise ratio as high as $\sim 10^3$ and noise
equivalent power below $10^{-19}$ W/$\sqrt{\text{Hz}}$ at 0.2 K have
been found to be achievable. Together with the available resolving power, the PJS is a promising candidate for single-photon detection in the THz regime.

We thanks the NanoSciERA "NanoFridge" project and the Academy
of Finland for financial support.


\begin{thebibliography}{99}
\bibitem{sinusoidalnote}
The expressions concerning the read-out have been derived using the
Josephson inductance model corresponding to the linearized
current-phase relation. This allows for analytic expressions, but
slightly underestimates the detector response.
\end{thebibliography}
\end{document}